\documentstyle[12pt]{article}
\begin{document}
\bibliographystyle{unsrt}
\newcommand{\bra}[1]{\left < \halfthin #1 \right |\halfthin}
\newcommand{\ket}[1]{\left | \halfthin #1 \halfthin \right >}
\newcommand{\be}{\begin{equation}}
\newcommand{\ee}{\end{equation}}
\newcommand{\vsig}{\mbox {\boldmath $\sigma$\unboldmath}}
\newcommand{\vep}{\mbox {\boldmath $\epsilon$\unboldmath}}
\newcommand{\fn}{\frac 1{E_N+M_N}}
\newcommand{\fs}{\frac 1{E_S+M_S}}

\title{\bf  The Kaon-Photoproduction Of Nucleons
In The Chiral Quark Model}

\author{Zhenping Li
\\
Physics Department, Carnegie-Mellon University \\
Pittsburgh, PA. 15213-3890 }
\maketitle

\begin{abstract}
In this paper, we develop a general framework to study the
meson-photoproductions of nucleons in the chiral quark model.  The S and U
channel resonance contributions are expressed in terms of
the Chew-Goldberger-Low-Nambu (CGLN)
amplitudes.  The kaon-photoproduction processes,  $\gamma p\to K^+ \Lambda$,
$\gamma p\to K^+ \Sigma^0$, and $\gamma p\to K^0\Sigma^+$, are
calculated.  The initial results show that the quark model provides a much
improved description  of the reaction mechanism for the
kaon-photoproductions of the nucleon with less parameters than the
traditional phenomenological approaches.
\end{abstract}

\newpage
\subsection*{\bf 1. Introduction}

The meson photoproduction of nucleons has always been a very important
field to study the structure of hadrons.   It was the investigations
by Copley, Karl and Obryk\cite{cko}, and Feynman, Kisslinger and
Ravndal\cite{fkr} in the pion-photoproduction 26 years ago that presented
first evidences of underlying $SU(6)\otimes O(3)$ symmetry for the baryon
structure in the quark model.  The interest in this field has been
recently revived by the establishment of the new experimental facilities
such as CEBAF.  In this paper, we present a new framework based on the
chiral quark model to study the meson-photoproductions of nucleons, in
particular the Kaon photoproductions.  Some of the highlights of this
project are
\begin{enumerate}
\item The transition amplitudes are chiral invariant in the low energy
limit.  Thus, the low energy theorem\cite{cgln} in the threshold pion
photoproduction is automatically recovered\cite{zpli94}.
\item It provides an unified description for all s- and u- channel resonance
contributions with only one set of parameters, which includes the
constituent quark masses and the coupling constant between the Goldstone
bosons, such as K, $\eta$ and $\pi$, and quarks.
\item  The contributions from the s and u channel resonances are expressed
in terms of the CGLN\cite{cgln} amplitudes so that the differential cross
section and various polarizations in meson photoproductions can be
easily carried out as the kinematics for the CGLN amplitudes is well
known\cite{tabakin}.
\item The form factors generated by the spatial integral lead to small
cross sections in high energy limit, which is consistent with the data.
This has been a major problem in the traditional hadronic
models\cite{thom,as90}.
\item It provides much improved description for the processes
$\gamma p \to K^+\Sigma^0$ and $\gamma p\to K^0 \Sigma^+$, while the
results in hadronic models\cite{mart} show a factor 10 to 100 larger
cross section for $\gamma p\to K^0\Sigma^+$ than that for $\gamma p\to
K^+\Sigma^0$.
\end{enumerate}

This project is an extension of our early investigation on the threshold
pion photoproduction\cite{zpli94}, in which the long-standing
problem\cite{dt92} of
the low energy theorem in the threshold pion-photoproduction was clarified.
Indeed,
although the quark model does give good descriptions of the electromagnetic
and strong decays of baryon resonances,  it does not guarrantee the
successes in the meson photoproductions.  The key is that one has to
combine the phenomenological quark model with chiral
symmetry\cite{MANOHAR}.  Moreover, since a baryon is being
 treated as a three quark system in the quark model, the separation of the
center of mass motion from the internal motion is crucial to  recover the
low energy theorem, this has been discussed in detail in the Compton
scattering $\gamma N\to \gamma N$\cite{zpli93} and the threshold
pion-photoproduction\cite{zpli94}.

In section 2, we present the framework for the Kaon-photoproductions
starting from the QCD Langrangian in the low energy region by assuming the
Kaon
as a Goldstone boson.  Of course, there would be chiral symmetry breaking
in Kaon productions. We shall only limit ourself in the symmetry limit and
in the same time, treat the coupling constant as a free parameter.
In section 3, we will present the formula for the leading order Born
amplitudes that include the seagul term, the nucleon exchange in the
s-channel, the $\Lambda$ and $\Sigma$ exchange in the u-channel, and the
kaon and $K^*$ exchange in the t-channel.  In section 4, the u-channel
contributions from the excited strange baryons are presented, in
particular, the $\Sigma^*$ resonances.   In section 5, we show how to write
the contributions from the nonstrange baryon resonances in terms of the
CGLN amplitudes.
In section 6, we present our numerical  evaluations of the differential
cross section, polarizations, and the total cross section for $\gamma p \to
K^+\Lambda$, $\gamma p\to K^+\Sigma^0$ and $\gamma p\to K^0\Sigma^+$.
Considering that  many effects have  not been included in our calculation and
no serious effort is being made to fit the experimental data,
the success of our approach is indeed very encouraging.
In section 7, we shall discuss some effects that have been left in our
calculation, and are important for the future investigations.
Although only the kaon transition amplitudes are presented
here, it should be pointed out that this approach is more general;
the $\eta$ and $\pi$ photoproductions of the nucleon can be related to the
Kaon photoproduction by the $SU(3)$ coefficients in each term. It is our
hope that this approach could provide a unified description of the
meson photoproductions of nucleons.

\subsection*{\bf 2. General Framework}
In the pure chiral symmetry limit, the low energy QCD Langrangian can be
written as\cite{MANOHAR}
\begin{equation}\label{i}
{\cal L}={\bar \psi} \left [ \gamma_{\mu} (i\partial^{\mu}+ V^\mu+A^\mu)
-m\right ] \psi + \dots
\end{equation}
where the vector and axial currents are
\begin{eqnarray}\label{ii}
V_\mu=\frac 12\left (\xi^\dagger
\partial_\mu\xi+\xi\partial_\mu\xi^\dagger\right ) ,
\nonumber \\
A_\mu=i\frac 12\left (\xi^\dagger \partial_{\mu} \xi -\xi\partial_{
\mu}
\xi^\dagger\right ),
\nonumber \\
\xi=e^{i\pi/f}
\end{eqnarray}
$f$ is a decay constant, and the field $\pi$ can be written in terms of
$SU(3)\otimes SU(3)$ Goldstone boson field.  The gauge transformation
of the axial vector $A_{\mu}$ in Eq. \ref{ii} leads to
a quark-photon-kaon vertex;
\begin{equation}\label{iii}
H_{K,e}=\sum_j \frac {e_K}{f_K}\phi_K {\bar \psi}_{j}(s)
\gamma_{\mu}^j\gamma_{5}^j \psi_j(u)A^{\mu}({\bf k},{\bf r}_j),
\end{equation}
where $A^{\mu}({\bf k},{\bf r}_j)$ is the electromagnetic field,
which generates a seagull term for the charge kaon production.  The
kaon-quark coupling at the tree level is therefore the standard
pseudovector coupling
\begin{equation}\label{iv}
H_K=\sum_j \frac 1{f_K} {\bar \psi}_j(s)\gamma_\mu^j\gamma_5^j \psi_j(u)
\partial^{\mu}\phi_K
\end{equation}
and the electromagnetic coupling is
\begin{equation}\label{v}
H_e=-\sum_j e_j \gamma^j_\mu A^\mu ({\bf k}, {\bf r}).
\end{equation}

Generally, one can write the transition matrix element as
\begin{eqnarray}\label{8}
{\cal M}_{fi}=\langle N_f| H_{K,e}|N_i \rangle +
\sum_j\bigg \{ \frac {\langle N_f|H_K |N_j\rangle
\langle N_j |H_{e}|N_i\rangle }{E_i+\omega-E_j} \nonumber \\
 +\frac {\langle N_f|H_{e}|N_j\rangle \langle N_j|H_K
|N_i\rangle }{E_i-\omega_K-E_j}\bigg \}+{\cal M}_T
\end{eqnarray}
where $N_i(N_f)$ is the initial (final) state of the nucleon,
and $\omega (\omega_{\pi})$ represents the
energy of incoming (outgoing) photons(pions).  The first
term in Eq. \ref{8} corresponds to the seagull diagram, which is a direct
consequence of the chiral transformations, the second term corresponds to
the S-channel resonance contribution which will be discussed in detail in
section 5, the third is the U-channel resonance contributions; they come
from the strange baryon resonances for the kaon photoproduction, and the
last term ${\cal M}_T$ is the T-channel meson exchange contribution. As
the couplings
between the excited strange mesons and the nucleon are not well known, we
will treat them as the free parameters, and in addition to the kaon
exchange, only the $K^*$ exchange will be included since the inclusion
of the other excited meson exchanges leads additional free parameters to
fit to the data.

The nonrelativistic expansion of the quark-photon-kaon
interaction gives
\begin{equation}\label{9}
H_{K,e}^{nr}=i\sum_j \frac e{f_K}
a^\dagger_j(s)a_j(u)\vsig_j \cdot \vep,
\end{equation}
where $\vep$ is the polarization vector of photons, where $a^\dagger_j(s)$
and $a_j(u)$ is the creation and annihilation operator for the strange and up
quarks for the charged kaon, while this term vanishes for the $K^0$
productions in the symmetry limit.
 Note that
\begin{equation}\label{10}
\langle N_f| \sum_j a^\dagger_j(s)a_j(u) {\vsig_j}
|N_i\rangle =g_A\langle N_f|{\vsig}|N_i\rangle ,
\end{equation}
where  $\vsig$ are the total spin operators of the nucleon, and $g_A$ is the
axial coupling constant for the kaon-nucleon-strange baryon couplings.  In
the $SU(6)\otimes O(3)$ symmetry limit, we have
\begin{equation}\label{11}
g_A=\left \{ \begin{array}{r@{\quad}l} \sqrt{\frac 32} & \mbox{ for
$KP\Lambda$} \\ -\frac 1{3\sqrt{2}} & \mbox{ for $KP\Sigma^0$}
\end{array}\right.  ,
\end{equation}
this leads to $1$ and $-\frac 13$ ratios between the couplings of axial
vector and vector for the $\Lambda$ and $\Sigma$ states.
Thus, we have an expression for the seagull diagram
\begin{equation}\label{12}
\langle N_f|H_{K,e}^{nr}|N_i\rangle =\frac {g_Ae}{f_K}
\langle N_f|{\vsig\cdot \vep}
 |N_i\rangle F({\bf k},{\bf q}) .
\end{equation}
where $F({\bf k}, {\bf q})$ is the form factor and the function of the
incoming photon and outgoing kaon momenta ${\bf k}$ and ${\bf q}$.

The differential cross section for the kaon-photoproduction in the center
of mass frame  is
\begin{equation}\label{13}
\frac {d\sigma^{c.m.}}{d\Omega}=\frac {\alpha_e \alpha_K(E_N+M_N)(E_S+M_S)}
{4s(M_S+M_N)^2}\frac {|{\bf q}|}{|{\bf k}|} |{\cal M^\prime}_{fi}|^2
\end{equation}
where the factor $eg_A/f_K$ has been removed from the transition matrix
elements ${\cal M^\prime}_{fi}$ so that it becomes dimensionless, and
$\sqrt {s}=E_N+\omega_\gamma=E_S+\omega_K$ is the total energy in the c.m.
frame.
The coupling constant $\alpha_K$ is related to the factor $g_A/f_K$ by the
generalized Goldberg-Treiman relation\cite{cad90}, however, the quark mass
effects lead to about 30 percent deviation from the measured value, while
the Goldberg-Trieman relation is accurate within 5 percent for the
pion couplings\cite{holstein}. Therefore, the coupling $\alpha_K$ will be
treated as a free parameter at present stage.

Generally, it is more convenient to study the
meson-photoproductions of
nucleon  in terms of the CGLN amplitude\cite{cgln};
\begin{equation}\label{14}
{\cal M^\prime}_{fi}={\bf J \cdot \vep}
\end{equation}
where $\vep$ is the polarization vector, and the current $J$ is
written as
\begin{equation}\label{15}
{\bf J}=f_1 \vsig+ if_2 \frac {(\vsig \cdot {\bf q})({\bf k}\times \vsig)}
{|{\bf q}| |{\bf k}|}+f_3\frac {\vsig \cdot {\bf k}}{|{\bf q}||{\bf k}|
}{\bf q}+f_4\frac {\vsig \cdot {\bf q}}{{\bf q}^2}{\bf q}
\end{equation}
in the center mass frame.  The differential cross section in terms of the
CGLN amplitude is\cite{tabakin}
\begin{eqnarray}\label{16}
|{\cal M}^\prime_{fi}|^2= & Re &\bigg \{ |f_1|^2+|f_2|^2-2\cos(\theta)
f_2f_1^*\nonumber
\\ & + & \frac
{\sin^2(\theta)}2 \left [
|f_3|^2+|f_4|^2+2f_4f^*_1+2f_3f_2^*+2\cos(\theta)f_4f_3^*\right ]\bigg \}
\end{eqnarray}
where $\theta$ is the angle between the incoming photon momentum ${\bf k}$
and outgoing kaon momentum ${\bf q}$ in the center mass frame.  The various
polarization observables can also be expressed in terms of CGLN amplitudes,
which can be found in Ref. \cite{tabakin}.

The calculation of the S- and U- channel resonances contributions follows a
procedure similar to that used in  Compton scattering
($\gamma N\to \gamma N$)\cite{zpli93}.  However, since our
investigation is not
limited to the low energy region, the relativistic kinematics should be
used for the  transition operator corresponding the center of mass motion;
 for example, the nonrelativistic
operator $\frac {{\bf P}}{2M_T}$ should be replaced by $\frac {{\bf
 P}}{E+M_T}$
where ${\bf P}$ and $E$ are the momentum and energy of the initial nucleon
or the final strange baryon.
Replacing the spinor $\bar \psi$ by $\psi^\dagger$ so
that the $\gamma$ matrices are replaced by the matrix {\boldmath
$\alpha$ \unboldmath}, the matrix elements for
the electromagnetic interaction $H_{e}$ can be written as
\begin{eqnarray}\label{17}
\langle N_j|H_e |N_i\rangle & = &\langle N_j|\sum_j e_j \mbox {\boldmath
$\alpha$ \unboldmath}_j \cdot \vep e^{i{\bf k}\cdot {\bf r}_j}  |N_i\rangle
\nonumber \\
& = & i \langle N_j|[\hat H, \sum_j e_j {\bf r}_j \cdot
\vep e^{i{\bf k}\cdot {\bf r}_j}] -\sum_j e_j {\bf r}_j\cdot \vep
\mbox {\boldmath $\alpha$ \unboldmath}_j\cdot \hat {\bf k}
e^{i{\bf k}\cdot {\bf r}_j}|N_i\rangle \nonumber \\
& = & i(E_j-E_i-\omega) \langle N_j| g_e|N_i\rangle +i\omega \langle N_j
|h_{e}|N_i\rangle ,
\end{eqnarray}
where
\begin{equation}\label{18}
\hat H= \sum_j (\mbox {\boldmath $\alpha$ \unboldmath}_j \cdot {\bf p}_j
+\beta_jm_j)+\sum_{i,j}V({\bf r}_i-{\bf r_j})
\end{equation}
is the Hamiltonian for the composite system,
\begin{equation}\label{19}
g_e=\sum_j e_j {\bf r}_j \cdot \vep e^{i{\bf k}\cdot {\bf r}_j},
\end{equation}
\begin{equation}\label{20}
h_e=\sum_j e_j {\bf r}_j\cdot \vep (1-\mbox {\boldmath
$\alpha$ \unboldmath}_j
\cdot \hat {\bf k})
e^{i{\bf k}\cdot {\bf r}_j},
\end{equation}
and $\hat {\bf k}=\frac {{\bf k}}{\omega_\gamma}$.
Similarly, we have
\begin{equation}\label{21}
\langle N_f|H_e |N_j\rangle
 = i(E_f-E_j-\omega) \langle N_f| g_e|N_j\rangle +i\omega_\gamma \langle N_f
|h_{e}|N_j\rangle .
\end{equation}
Therefore the second and the third terms in Eq. \ref{8} can be written
as
\begin{eqnarray}\label{22}
{\cal M}_{23}^{\prime}=i \langle N_f|[g_e,H_K]|N_i\rangle
+ i\omega \sum_j\bigg \{ \frac {\langle N_f|H_K
|N_j\rangle \langle N_j |h_{e}|N_i\rangle }{E_i+\omega-E_j}
\nonumber \\  +\frac {\langle N_f|h_{e}|N_j
\rangle \langle N_j|H_K|N_i\rangle }{E_i-
\omega_K-E_j}\bigg \}.
\end{eqnarray}
The nonrelativistic expansion for $h_e$ in Eq. \ref{22}
is\cite{zpli93,zpli94}
\begin{equation}\label{23}
h_e=\sum_j \left [ e_j {\bf r}_j \cdot \vep (1-\frac {{\bf p}_j \cdot
{\bf k}}{m_j \omega})-\frac {e_j}{2m_j}{\vsig_j \cdot (\vep\times \hat
 {\bf k})} \right ],
\end{equation}
which $h_e$ is only expanded to order $1/m$, and it has been
shown\cite{zpli94} that the expansion to order $1/m_q$ is sufficient to
reproduce the low energy theorem for the threshold
pion-photoproductions\cite{cgln}.  The corresponding kaon-coupling  for the
initial nucleon and final $\Lambda$ or $\Sigma$ states is being
written as
\begin{eqnarray}\label{24}
H_{K}^{nr} =\sum_j \bigg \{ \frac {\omega_K}{E_S+M_S} \vsig_j\cdot
{\bf P}_f+\frac {\omega_K}{E_N+M_S}\vsig_j \cdot {\bf P}_i-
\vsig_j\cdot {\bf q}\nonumber \\ +\frac {\omega_K}{2\mu_s}\vsig_j
\cdot {\bf p}_j \bigg \}\frac {a^\dagger_j(s)a_j(u)}{g_A}
\end{eqnarray}
where $\omega_K$ is the energy of the emitting kaons, and $\frac
1{\mu_s}=\frac 1{m_s}+\frac 1{m_q}$.  The first three terms
in Eq. \ref{24} corresponds to the center of mass motion, and the last
term represents the internal transition.  Similarly, the $K^*$ coupling
between the nucleon and strange baryon has the structure;
\begin{eqnarray}\label{241}
H_{K^*}^{nr} =g_{K^*}\sum_j \bigg \{ \frac 1{E_S+M_S} P_S\cdot \epsilon_v
+\frac 1{E_N+M_S}P_N\cdot \epsilon_v\nonumber \\
+\frac {1}{2m_j}{\vsig_j
\cdot (\vep_v\times \hat
 {\bf k})} \bigg \}\frac {a^\dagger_j(s)a_j(u)}{g_A},
\end{eqnarray}
where $P_S\cdot \epsilon_v=E_S\epsilon_v^0-{\bf P}_S\cdot \vep_v$,
and only the operator associated with the c.m. motion is presented.

\subsection*{\bf 3. The Leading Order Born Amplitudes}

Following the same procedure in Ref \cite{zpli94},  the amplitudes
for the seagull term is
\begin{equation}\label{25}
{\cal M}_s=F({\bf k},{\bf q}) \left [ 1+\frac {\omega_K}2 \left
(\fn+\fs\right )\right ]\vsig\cdot\vep,
\end{equation}
thus, the the seagull term only contribute to the CGLN amplitude $f_1$.
If the outgoing kaon is being considered as point like particle, the form
factor would be
\begin{equation}\label{26}
F({\bf k},{\bf q})=exp\left ( -\frac {({\bf k}-{\bf
q})^2}{6 \alpha^2}\right )
\end{equation}
in the harmonic oscillator basis, where $\alpha$ is the oscillator
strength.   One could use the
quark pair creation model\cite{LY} to calculation this form factor for
the finite size kaons, we find that Eq. \ref{26} would be modified by the
additional form factor that are function of ${\bf k}^2$ and ${\bf q}^2$,
in the same time, the finite size kaon also destroys the chiral
symmetry (see discussion in section 7).

The matrix element for the nucleon pole term is found to be
\begin{eqnarray}\label{27}
{\cal M}_N= - \omega_K e^{-\frac {{\bf q}^2+{\bf k}^2}{6\alpha^2}}
\left ( \fs+\fn\right )\left (1+\frac {{\bf k}^2}{4P_N\cdot k}\mu_N \right )
 \vsig\cdot\vep \nonumber \\  +   ie^{-\frac {{\bf k}^2+{\bf q}^2}
{6\alpha^2}}\left [ \frac {\omega_K}2\left (\fs+\fn\right)+1\right ]\frac
{\mu_N}{2P_N\cdot k} \vsig
\cdot {\bf q} \vsig \cdot (\vep \times {\bf k})
\end{eqnarray}
where $P_N\cdot k=\omega_{\gamma}(E_N+\omega_{\gamma})$,
 and $\mu_N$ is the magnetic moments of the nucleon.

The matrix elements for
the U-channel $\Lambda$ and $\Sigma$ exchange term is
\begin{eqnarray}\label{28}
{\cal M}_{\Lambda\Sigma}=-e^{-\frac {{\bf k}^2+{\bf q}^2}{6\alpha^2}}
\frac {M_S}{2M_N}\left (\frac {\mu_S}{P_S\cdot k}+\frac
{g_{\Lambda\Sigma}\mu_{\Lambda\Sigma}}{P_S\cdot k\pm \delta m^2}\right )
\nonumber \\
\bigg \{ \frac {\omega_K{\bf k}^2}2 \left ( \fs+\fn\right ) \vsig \cdot
\vep -\nonumber \\ i \left [ \frac {\omega_K}2 \left ( \fs+\fn\right )
+1\right ] \vsig \cdot
(\vep\times {\bf k}) \vsig \cdot {\bf q}\bigg \}
\end{eqnarray}
where
\begin{equation}\label{29}
g_{\Lambda\Sigma}=\left \{ \begin{array}{r@{\quad}l}
\frac {g_\Sigma}{g_\Lambda} & \mbox {for $\gamma p\to K^+\Lambda$} \\
\frac {g_\Lambda}{g_\Sigma} & \mbox {for $\gamma p\to K^+\Sigma^0$} \\
0 & \mbox{ for $\gamma p\to K^0 \Sigma^+$} \end{array} \right.
\end{equation}
is the ratio between the coupling constants for $\Lambda$ and $\Sigma$
final states, $\mu_{\Lambda\Sigma}=1.61$ is the magnetic moments for the
transition between the $\Lambda$ and $\Sigma^0$ states, and $P_S\cdot k=
E_S\omega_{\gamma}+{\bf k}\cdot {\bf q}$, notice that the final baryon
state has the total momentum $-{\bf q}$ in the center of mass system.
If the $\Lambda$ and $\Sigma$ pole terms in the U-channel are replaced
by the nucleon,  and kaons replaced by pions,  the low energy theorem
 in the threshold pion-photoproductions should be recovered by combining
Eqs.  \ref{24}, \ref{27} and \ref{28}.

The matrix elements for the t-channel are
\begin{equation}\label{30}
{\cal M}_K=e^{-\frac {{\bf k}^2}{8\beta^2}-\frac {({\bf k}-{\bf
q})^2}{6\alpha^2}}\frac {(M_S+M_N){\bf q}\cdot \vep}{q\cdot k}\left (\fs\vsig
 \cdot {\bf q}-\fn \vsig \cdot {\bf k}\right )
\end{equation}
for the kaon exchange and
\begin{eqnarray}\label{31}
{\cal M}_{K^*}=e^{-\frac {{\bf k}^2}{8\beta^2}-\frac {({\bf k}-{\bf
q})^2}{6\alpha^2}}\frac {(M_{K^0}+M_{K^*})g_{K^*}}{t-M_{K^*}^2}
\bigg \{ \frac 1{2\mu_s} \vsig\cdot \left (({\bf q}-{\bf k})
\times ({\bf k}\times \vep)\right ) \nonumber \\
+ i \frac {g_V}{g_A} {\bf q}\cdot ({\bf k}\times \vep)\left [
\fs\left (1+\frac {M_S^2-M_N^2+t}{2M_{K^*}^2}\right )-\frac
 {t+M_N^2-M_S^2}{2M_{K^*}^2(E_N+M_N)}\right ]\bigg\}
\end{eqnarray}
for the $K^*$ exchange, where $t=M_K^2-2(\omega_{\gamma}\omega_K-{\bf
q}\cdot {\bf k})$. The parameter
$g_{K^*}$ in Eq. \ref{31} is the ratio between the couplings of $K^0$
 and $K^*$ to the nucleon and the strange baryon.

\subsection*{\bf 4. The U-channel resonance contribution}

The leading U-channel contribution comes from the resonance
$\Sigma^*(1385)$.  In the quark model, it belongs to the same ${\bf 56}$
multiplet as the resonances $\Lambda$ and $\Sigma$, and there should be no
orbital excitations in the symmetry limit.  Thus, only the c.m. motion part
of transition operator in Eq. \ref{24} would contribute.  We can rewrite
the transition operator corresponding to the c.m. mass motion in Eq.
\ref{24} as
\begin{equation}\label{32}
H_{K}^c=\sum_j \vsig_j\cdot {\bf A}\frac {a^\dagger_j(s)a_j(u)}{g_A}
\end{equation}
where
\begin{equation}\label{33}
{\bf A}=-\omega_K\left (\fn +\fs\right ){\bf k}-\left ( \omega_K\fs
+1\right ){\bf q}.
\end{equation}
Notice that the initial nucleon and intermediate $\Sigma$ states have the
c.m. momenta $-{\bf k}$ and $-{\bf k}-{\bf q}$ respectively.  Then the
contribution from the $\Sigma(1385)$ can be written as
\begin{equation}\label{34}
{\cal M}_{\Sigma(1385)}=\frac {M_Sg_Se^{-\frac {{\bf q}^2+{\bf
k}^2}{6\alpha^2}}}{M_N(P_S\cdot k+\delta M^2_{\Sigma^*}/2)}\left [i2\vsig \cdot
{\bf A} \vsig\cdot
(\vep\times {\bf k})+\vsig\cdot ({\bf A}\times (\vep\times {\bf k}))\right ]
\end{equation}
where the factor $g_S$ is
\begin{equation}\label{35}
g_S=\left \{ \begin{array}{r@{\quad}l}
-\frac {\mu_{\Lambda}}{3} & \mbox{ for $\gamma p\to K^+\Lambda$} \\
{4\mu_{\Sigma^0}} & \mbox{ for $\gamma p\to K^+\Sigma^0$}  \\
\frac {2\mu_{\Sigma^+}}{3} & \mbox{ for $\gamma p\to K^0\Sigma^+$}
\end{array} \right. ,
\end{equation}
and $\delta  M^2_{\Sigma^*}=M^2_{\Sigma^*}-M^2_S$.  The magnetic
moments in Eq. \ref{35} are $\mu_{\Lambda}=-0.61$, $\mu_{\Sigma^0} \approx 1.0$
and $\mu_{\Sigma^+}=2.38$.

For the excited resonances, we follow the procedure in Ref.
\cite{zpli93}.  In the harmonic oscillator basis, one can write the
transition matrix elements for the excited strange baryon resonances as
\begin{equation}\label{36}
{\cal M}_Y=\left ({\cal M}^3_Y+{\cal M}^2_Y\right )
e^{-\frac {{\bf k}^2+{\bf q}^2}{6\alpha^2}}
\end{equation}
in the symmetry limit.  The first term represents
the process in which the incoming photon and outgoing kaon are absorbed and
emitted by the same quarks, its general
 expression can be obtained by the
second quantization approach in the harmonic oscillator basis\cite{zpli93};
\begin{eqnarray}\label{37}
{\cal M}^3_Y= \frac {-1}{6m_s}
\left [i\frac {g_V}{g_A} {\bf A}\cdot (\vep\times {\bf k})+\vsig\cdot ({\bf
A}\times (\vep\times {\bf k}))\right ]F(\frac {{\bf k}\cdot {\bf q}}{3
\alpha^2}, -P_S\cdot k) \nonumber \\
+\frac 1{18}\left [\frac {\omega_K\omega_{\gamma}}{\mu_s}\left (1+\frac
{\omega_{\gamma}}{2m_s}\right )\vsig \cdot \vep+\frac
{2}{\alpha^2}\vsig\cdot {\bf A}\vep\cdot {\bf q}\right ]
  F(\frac {{\bf k}\cdot {\bf q}}{3\alpha^2}, -P_S\cdot k-\delta M^2)
\nonumber \\ +\frac {\omega_K\omega_r}{54\alpha^2\mu_s}
  F(\frac {{\bf k}\cdot {\bf q}}{3
\alpha^2},-P_S\cdot k-2\delta M^2).
\end{eqnarray}
The function $F(x, y)$ in Eq. \ref{37} corresponds to the product of the
spatial integral and the propagator for the excited states, it can be
written as
\begin{equation}\label{377}
F(x, y)=\sum_n \frac {M_S}{ n! (y-n\delta M^2)} x^n,
\end{equation}
where $n\delta M^2=(M_n^2-M^2)/2$ represents the mass
difference between the ground state and excited states with the major
quantum number $n$ in the harmonic oscillator basis, which will be chosen as
the average mass differences between the ground state and the negative
parity baryons.  Therefore, first
term in Eq. \ref{37} corresponds to the correlation between the magnetic
transition and the c.m. motion of the kaon transition operator, it
contributes to the leading Born terms in the U-channel. The second term
in Eq. \ref{37} is the correlations among the internal and c.m. motions of
the photon and kaon transition operators, this term only contributes to the
transitions between the ground and $n\ge 1$ excited states in the
harmonic oscillator basis.
The third term in Eq. \ref{37} corresponds to
the correlation of the internal motions between the photon and kaon
transition operators, which only contributes to the transition between the
ground and $n\ge 2$ excited states.
In the $SU(6)$ symmetry limit, the ratio $\frac {g_V}{g_A}$ is $1$ for the
$\Lambda$ final state and $-3$ for the $\Sigma$ final state.
The second term ${\cal M}_2$ in Eq. \ref{36} represents the
process in which the  incoming photon
and outgoing kaon are absorbed and emitted by different quarks,  it can be
written as
\begin{eqnarray}\label{38}
\frac {{\cal M}^2_Y}{g_S^\prime}
= \frac 1{6m_q} \left [ -ig_v^\prime {\bf A}\cdot (\vep \times
{\bf k})+g_a^\prime \vsig \cdot ({\bf A}\times (\vep\times {\bf k}))\right ]
 F(\frac {-{\bf k}\cdot {\bf q}}{6\alpha^2}, -P_S\cdot k)\nonumber
\\
-\frac 1{36}\left [\frac {\omega_K\omega_{\gamma}}{\mu_s}
\left (1+g_a^\prime\frac {\omega_{\gamma}}{2m_q}\right )+\frac
{2\omega_{\gamma}}{\alpha^2} \vsig \cdot {\bf A}\vep\cdot {\bf q}\right ]
F(-\frac {{\bf k}\cdot {\bf q}}{6
\alpha^2},-P_S\cdot k-\delta M^2) \nonumber  \\
\frac {\omega_K\omega_\gamma}{216\alpha^2 \mu_s}\vsig \cdot {\bf
k}\vep\cdot {\bf q} F(-\frac {{\bf k}\cdot {\bf q}}{6\alpha^2},-P_S\cdot
k-2\delta M^2)
\end{eqnarray}
where the factor $g_S^\prime$ equals to $1$ for both charged kaon
productions, 4 for the $K^0$ productions, the factors $g_v^\prime$ and
$g_a^\prime$ are
\begin{equation}\label{39}
g_v^\prime=\left \{ \begin{array} {r@{\quad}l} 1 & \mbox{for $\gamma p\to
K^+\Lambda$} \\ -7 & \mbox{for $\gamma p\to K^+\Sigma^0$} \\
1 & \mbox{for $\gamma p\to K^0\Sigma^+$} \end{array} \right.
\end{equation}
and
\begin{equation}\label{40}
g_a^\prime=\left \{ \begin{array} {r@{\quad}l} 1 & \mbox{for $\gamma p\to
K^+\Lambda$} \\ 9 & \mbox{for $\gamma p\to K^+\Sigma^0$} \\
0 & \mbox{for $\gamma p\to K^0\Sigma^+$} \end{array} \right.  .
\end{equation}
An interesting observation from these expressions is that  the transition
matrix elements ${\cal M}_Y^3$ and ${\cal M}^2_Y$ that correspond to the
incoming photons and outgoing kaons being absorbed and emitted by the same
and different quarks differ by a factor $\left (-\frac 12\right )^n$,  thus
the transition matrix element ${\cal M}^3_Y$ becomes dominant as the
quantum number $n$ increases.

Eqs. \ref{36} and \ref{38} can be summed up to any quantum number $n$,
however, the excited states with large quantum number $n$ become
less significant for the U-channel resonance contributions.  Thus, we only
include the excited states with $n\le 2$, which is the minimum number
required for the contribution from every term in Eqs. \ref{36} and
\ref{38}.

\subsection*{\bf 5. The S-channel resonance contribution}

There are two major components for the S-channel resonance contributions;
the well known baryon resonances below $2$ GeV that correspond to $n\le 2$
according to the $SU(6)\otimes O(3)$ classification and the resonances
above 2 GeV that are not well known both theoretically and experimentally.
One could regard the resonances above 2 GeV as the continuum
contributions, however, they must be included in the calculations if one
intends to calculate the photoproductions above 2 GeV in the c.m. frame.
The advantage of the quark model approach is that it
provides us a unified description of the S- and U-channel
resonance contributions and the continuum contributions by the same
set of parameters.

For the S-channel resonance processes, the operator ${\bf A}$ in Eq.
\ref{32} should be
\begin{equation}\label{41}
{\bf A}=-\left (\omega_K\fs+1\right ) {\bf q}
\end{equation}
in the c.m. frame.  The calculation of the S-channel resonance
contributions is similar to that of the U-channel resonance contributions.
However, since the operator ${\bf A}$
is only proportional to the final state momentum ${\bf q}$, the partial
wave analysis can be easily carried out for the S-channel resonances.

In general, one can write the S-channel resonance amplitudes as
\begin{equation}\label{42}
{\cal M}_R=\frac 2{s-M_R^2}e^{-\frac {{\bf k}^2+{\bf q}^2}{6\alpha^2}}
{\cal O}_R,
\end{equation}
where $\sqrt {s}=E_N+\omega_{\gamma}=E_S+\omega_K$
is the total energy of the system, and ${\cal O}_R$ is determined
by the structure of each resonance.  Eq. \ref{42} shows that there should be
a form factor,  $e^{-\frac {{\bf k}^2+{\bf
q}^2}{6\alpha^2}}$ in the harmonic oscillator basis, even in the real
photon limit.  This leads to a smaller cross section in the high energy
limit, which has been a major problem in the hadronic
models\cite{as90,thom}.  If the mass of a resonance is above the
threshold, the mass $M_R$ in Eq. \ref{42}
should be changed to
\begin{equation}\label{43}
M_R^2 \to M_R(M_R-i\Gamma({\bf q})).
\end{equation}
$\Gamma({\bf q})$ in Eq. \ref{43} is the total width of the resonance,
and a function of the final state momentum ${\bf q}$.  For a resonance
decay to a two body final state with orbital angular momentum $l$,
the decay width $\Gamma({\bf q})$ can be written as
\begin{equation}\label{44}
\Gamma({\bf q})= \Gamma_R \left (\frac {|{\bf q}|}{|{\bf q}_R|}\right )^{2l+1}
\frac {\sqrt {s}}{M_R} \frac {D_l({\bf q})}{D_l({\bf q}_R)},
\end{equation}
where $|{\bf q}_R|=\sqrt{\frac {(M_R^2-M_S^2+M_K^2)^2}{4M_R^2}-M_K^2}$
and $\Gamma_R$ are the
momentum of one final state and the decay width
at the S-channel resonance mass $M_R$.  The
function $D_l({\bf q})$ is called fission barrier\cite{bw}, and wavefunction
dependent; here we use
\begin{equation}\label{45}
D_l({\bf q})=exp\left (-\frac {{\bf q}^2}{3\alpha^2}\right ),
\end{equation}
which is independent of $l$.
A similar formula used in I=1 $\pi\pi$ and and p-wave $I=1/2$
$K\pi$ scattering was found in excellent
agreement with data in the $\rho$ and $K^*$ meson region\cite{barnes}.
 Of course,
whether Eq. \ref{45} is suitable for the baryon resonance is unclear, and
should be studied in the $P_{33}(1232)$ region of pion-photoproduction,
where the data are well known.

The first S-channel resonance is the resonance $P_{33}(1232)$, the
transition matrix element is similar to that of the resonance
$\Sigma^*(1387)$ in U-channel, however, it does not contribute to the process
$\gamma p\to K^+\Lambda$ due to the isospin couplings.  The matrix element
for the $\Sigma$ final state is
\begin{equation}\label{46}
{\cal O}_{P_{33}(1232)}=\mu_Ng_{\Delta}\left [ i2{\bf A}\cdot (\vep \times
{\bf k})+\vsig \cdot ({\bf A}\times (\vep\times {\bf k}))\right ],
\end{equation}
where
\begin{equation}\label{47}
g_{\Delta}=\left \{ \begin{array}{r@{\quad}l}
 \frac 43 & \mbox{ for $\gamma p\to
K^+\Sigma^0$} \\ -\frac 23 & \mbox{ for $\gamma p\to K^0\Sigma^+$}
\end{array}\right.
\end{equation}
is determined by the $SU(6)$ symmetry. Any isospin 3/2 state in ${\bf 56}$
representation should have the same factor $g_{\Delta}$.
It is straightforward to express the transition matrix element in Eq.
\ref{46} in terms of the CGLN amplitudes, and the multiple decomposition of
the CGLN amplitudes (see Ref. \cite{tabakin}) shows that the transition
matrix element in Eq. \ref{46} has the characteristics of the $M_1^+$
transition.

For the P-wave baryon resonance with the quantum number $n=1$, there are
seven resonances which have been identified experimentally.  However,
the photon transitions from the proton target are simplified due to the
Moorhouse selection rule\cite{moor};  the photoabsorption amplitudes for
$\gamma p\to N^*$  belonging to $SU(3)$ octets with spin $\frac 32$
vanish for the transition operator in Eq. \ref{23}.  This reduces the
number of the resonances that contribute to the kaon photoproduction of the
proton target to 4.   Furthermore,  the isospin couplings for $N^*\to K^+
\Lambda$ only allow the resonances with isospin $\frac 12$, which reduces
the number of the P-wave resonances to 2 for $\Lambda$ final state.
We find that the transition amplitude for the resonance with the
quantum number ${\bf 70},N(^2P_M){\frac 12}^-$ is
\begin{equation}\label{48}
{\cal O}_{S_{11}}=g_{M}\frac {\omega_{\gamma}}{18}\left (\frac
{3\omega_K}{2\mu_s}+\frac 1{\alpha^2}{\bf q}\cdot {\bf A}\right )\left (1
+\frac {|{\bf k}|}{2 m_q}\right )\vsig\cdot \vep
\end{equation}
where $g_{M}$ factor is $1$ for the $\Lambda$ final state
and $-1$ for both $\Sigma^0$ and $\Sigma^+$ final state.
This transition matrix element
 is a pure CGLN amplitude $f_1$, and independent of the scattering angle
since it is a S-wave.  It is also a product of the amplitudes for $\gamma
p\to S_{11}$ and kaon decays.  The photoabsorption amplitude agrees with
the result in Ref. \cite{zpli90}, and the kaon decay amplitude is the same
as the expression in Table 1 in Ref. \cite{simon} with $g-\frac 13h=\frac
{|{\bf A}|}{|{\bf q}|}$, and $h=\frac {\omega_K}{2\mu_s}$.  Note that
${\bf A}$ has a
negative sign,  this is consistent with the fitted value for $g-\frac 13h$
and $h$ in Ref. \cite{simon}.  There are two $S_{11}$ resonances
with masses around $1.6$ GeV; $S_{11}(1535)$ and $S_{11}(1650)$.
Traditionally, the resonance $S_{11}(1535)$ is being classified as a
${\bf 70},N(^2P_M){\frac 12}^-$ state, however, the potential quark model
calculation\cite{IK79} found that the two states, ${\bf 70}N(^2P_M){\frac
12}^-$ and ${\bf 70}N(^4P_M){\frac 12}^-$, are strongly mixed.  Thus, we
have
\begin{eqnarray}\label{49}
{\cal O}_{S_{11}(1535)} =0.6 {\cal O}_{S_{11}} \nonumber \\
{\cal O}_{S_{11}(1650)} =0.4 {\cal O}_{S_{11}}
\end{eqnarray}
where the coefficients $0.6$ and $0.4$ come from the mixing angle obtained
from the potential model calculation.

The transition amplitude for the resonance $D_{13}(1520)$ is
\begin{eqnarray}\label{50}
{\cal O}_{D_{13}(1520)}=-g_{M}\frac 1{12}
\bigg [ \frac {i{\bf q}\cdot
{\bf k}}{m_q\alpha^2}\vsig\cdot {\bf A}\vsig \cdot (\vep\times {\bf
k})+\frac 13 \left (\frac {{\bf k}^2}{m_q\alpha^2}+\frac
{2\omega_{\gamma}}{\alpha^2}\right ){\bf A}\cdot {\bf q}\vsig\cdot \vep
\nonumber \\ -
\frac {2\omega_{\gamma}}{\alpha^2} \vsig\cdot {\bf A}{\bf q}\cdot
\vep\bigg ]
\end{eqnarray}
where the factor $g_{M}$ is the same as that in Eq. \ref{48}, since the
states $S_{11}$ and $D_{13}$ belong to the same $SU(3)$ representation.
  Notice that
there is little mixing between the spin 1/2 and spin 3/2 state for the
resonance $D_{13}(1520)$\cite{IK79}.

If one treats the transition amplitudes ${\cal O}$ for the
resonance $S_{11}$ and $D_{13}$ in Eqs. \ref{48} and \ref{50} as a function
of the quark mass $m_q$, the transition amplitudes for the isospin 3/2
states can be related to those for the resonances $S_{11}$ and $D_{13}$.
The amplitudes for the resonances $S_{31}(1670)$ and $D_{33}(1700)$ are
\begin{eqnarray}\label{51}
{\cal O}_{S_{31}(1670)}=2{\cal O}_{S_{11}}(g_{M} \to g_{\Delta_{M}}, m_q\to
-3m_q), \nonumber \\
{\cal O}_{D_{33}(1700)}=2{\cal O}_{D_{13}}(g_{M} \to g_{\Delta_{M}}, m_q\to
-3m_q)
\end{eqnarray}
which replace the factor $g_{M}$ in Eqs. \ref{48} and \ref{50}
by $g_{\Delta_M}$, and the quark mass $m_q$ by $-3m_q$.
The factor $g_{\Delta_M}$ is 1 for the charged kaon
production and $-\frac 12$ for the $K^0$ production, and these two states
do not couple to the $\Lambda$ final state.

The situation for the positive parity baryon states with the quantum number
$n=2$ is not as clear as the P-wave baryons.  The well known resonances
belong to the ${\bf 56}$ multiplets.  The transition amplitudes for the
radial excitation states are
\begin{equation}\label{52}
{\cal O}_{P_{11}(1440)}=-\frac {{\bf k}^2}{216\alpha^2}
\left (\frac {{\bf A}\cdot {\bf q}}{\alpha^2}+\frac {\omega_K}{\mu_s}\right
) i\vsig \cdot {\bf q}\vsig\cdot (\vep\times {\bf k})
\end{equation}
for the Roper resonance, and
\begin{equation}\label{53}
{\cal O}_{P_{33}(1600)}=\frac {g_{\Delta}{\bf k}^2}{108m_q\alpha^2}
\left (\frac {{\bf A}\cdot {\bf q}}{\alpha^2}+\frac {\omega_K}{\mu_s}\right
) \bigg [i2{\bf q}\cdot (\vep\times {\bf k})
+\vsig\cdot ({\bf q}\times (\vep\times {\bf k})\bigg ]
\end{equation}
for the resonance $P_{33}(1600)$, where the factor $g_{\Delta}$ is given in
Eq. \ref{47}.

The transition  amplitudes for the $N(^2D_S)$ states are
\begin{eqnarray}\label{54}
{\cal O}_{P_{13}(1720)}=\frac {{\bf k}^2}
{90\alpha^2}
\left (\frac {{\bf A}\cdot {\bf q}}{\alpha^2}+\frac {5\omega_K}{2\mu_s}
\right ) \bigg [ \frac {i\omega_{\gamma}}{6m_q}\vsig\cdot {\bf q} \vsig
\cdot (\vep\times {\bf k})\nonumber \\
+\vsig\cdot \vep {\bf k}\cdot {\bf q} \left
(1+\frac {\omega_{\gamma}}{2m_q}\right )+\vsig\cdot {\bf k} {\bf q}\cdot
\vep\bigg ]
\end{eqnarray}
for the resonance $P_{13}(1720)$, and
\begin{eqnarray}\label{55}
{\cal O}_{F_{15}(1688)}=\frac 1{36\alpha^4}
\bigg \{
\omega^3_{\gamma}\left [ \vsig \cdot {\bf A} \vep\cdot {\bf q} {\bf k}\cdot
{\bf q}-\frac 15{\bf q}^2(\vsig \cdot \vep {\bf k}\cdot {\bf A}+\vsig \cdot
{\bf k}\vep\cdot {\bf A})\right ]\nonumber \\-
\frac 1{4m_q}\left [i\vsig \cdot {\bf A}\vsig \cdot (\vep\times
{\bf k})\left ( ({\bf k}\cdot {\bf q})^2-\frac 15 {\bf q}^2{\bf k}^2\right
)+\frac 25\vsig\cdot\vep ({\bf k}\cdot {\bf q})^2 {\bf k}\cdot {\bf A}
\right ]\bigg \}
\end{eqnarray}
for the resonance $F_{15}(1688)$.  The isospin 3/2 states $\Delta
(^4D_S)$ are concentrated around mass 1.9 GeV, therefore, we simply treat
them as degenerate.  The amplitude for the resonances $P_{31}(1910)$ and
$P_{33}(1920)$ for the $\Sigma$ final state gives
\begin{equation}\label{56}
{\cal O}_{P_{3(1+3)}}=
\frac {-g_{\Delta}{\bf k}^2}{108m_q\alpha^2}
\left (\frac {2{\bf A}\cdot {\bf q}}{5\alpha^2}+\frac {\omega_K}{\mu_s}\right
) \bigg [i2{\bf q}\cdot (\vep\times {\bf k})
+\vsig\cdot ({\bf q}\times (\vep\times {\bf k})-3\vsig\cdot\vep {\bf
k}\cdot {\bf q}\bigg ].
\end{equation}
The amplitude for the resonances $F_{35}(1905)$ and $F_{37}(1950)$
is
\begin{eqnarray}\label{57}
{\cal O}_{F_{3(5+7)}}=\frac {g_{\Delta}}{36m_q\alpha^4}\bigg [\left
(i2{\bf A}\cdot (\vep\times {\bf k})
+\vsig\cdot ({\bf A}\times (\vep\times {\bf k})\right )\nonumber \\
\left ( ({\bf k}\cdot {\bf q})^2-\frac 15 {\bf k}^2{\bf q}^2\right ) +
\frac 25 \vsig \cdot \vep {\bf k}\cdot {\bf A} {\bf k}^2{\bf q}^2\bigg ].
\end{eqnarray}
These are the transition amplitudes for the {\bf 56} multiplet.  The
coefficients for the isospin 3/2 states in the ${\bf 56}$ multiplet are
significantly larger than those for the isospin 1/2 states for the
$\Sigma$ final states.  One also finds that the contributions from the
Roper resonance $P_{11}(1440)$ are small compare to the resonance
$F_{15}(1688)$.  However, the traditional quark model does not give a good
description for the Roper resonances,  and the gluonic degrees of freedom may
play an explicit role in this resonance\cite{zpli91},  the consequences
from the gluonic degrees of freedom in the Roper resonance remain
 to be studied.

The contributions from those resonances belonging to the ${\bf 70}$
 multiplet can be related to the transition amplitudes of the ${\bf 56}$
multiplets.  The resonance $P_{11}(1710)$ which is being assigned as a
$N(^2S_{M}){\frac 12}^+$ state can be related to the Roper resonance by
\begin{equation}\label{58}
{\cal O}_{P_{11}(1710)}=g_{M}\frac 12 {\cal O}_{P_{11}(1440)}.
\end{equation}
  The contribution from the
resonance $P_{31}(1750)$ is
\begin{equation}\label{59}
{\cal O}_{P_{31}(1750)}=g_{\Delta_M}\frac 13 {\cal O}_{P_{11}(1440)}.
\end{equation}
 Indeed, the electromagnetic coupling of this resonance is very weak,
which was only seen in the $\pi N$ scattering\cite{manley}.  If we assume
the resonance $F_{15}(2000)$ as a ${\bf 70} N(^2D_M){\frac 52}^+$ state,
its transition
amplitude is related to that of the
resonance $F_{15}(1688)$ by
\begin{equation}\label{60}
{\cal O}_{F_{15}(2000)}=g_{M}\frac 12 {\cal O}_{F_{15}(1688)}.
\end{equation}
These relations are determined by the $SU(6)\otimes O(3)$ symmetry. One could
easily obtain the  transition amplitudes for other ${\bf 70}$ multiplet
states,  we  will neglect those resonance here since their couplings are very
weak from our calculation, which do not make much difference even if they
are included, and their width and masses have not been determined
experimentally.

The transition amplitudes from the S-channel resonances provide us some
important insights into the role of the baryon resonances in the kaon
photoproduction even without further numerical evaluation; the resonances of
the higher partial waves are very important in the process $\gamma p\to
K^+\Lambda$, in particular the resonance $D_{13}(1520)$ of P-wave baryons
and the resonance $F_{15}(1688)$ of $n=2$ baryon state,  which are
usually neglected in the traditional hadronic model.  For the
processes $\gamma p\to K^+\Sigma^0$ and $\gamma p\to K^0\Sigma^+$, the
contributions from the isospin 3/2 states, in particular those resonances
in ${\bf 56}$ multiplet, are dominant.  Therefore, the processes $\gamma
p\to K^+\Sigma^0$ and $\gamma p\to K^0\Sigma^+$  provide us a very
important probe to the resonances with isospin 3/2, a particular example is
the resonances $F_{37}(1950)$, $F_{35}(1905)$, $P_{33}(1920)$ and
$P_{31}(1910)$.

If one intends to calculate the reaction beyond 2 GeV in the center of mass
frame, the higher resonances with quantum number $n=3$ and $n=4$ must be
included.  There is a little knowledge of resonances in these region except
 high partial wave resonances.   However, we can assume that the
resonances for $n\ge 2$ are degenerate, the sum of the transition
amplitudes from these resonances can be obtained through the approach in
Ref. \cite{zpli93}.  The transition amplitude for the nth harmonic
oscillator shell is
\begin{equation}\label{61}
{\cal O}_{n}={\cal O}_n^2 +{\cal O}_n^3
\end{equation}
where the amplitudes ${\cal O}_n^2$ and ${\cal O}_n^3$ have the same
meaning as the amplitudes ${\cal M}_Y^2$ and ${\cal M}_Y^3$ in Eqs.
\ref{37} and \ref{38}, and we have
\begin{eqnarray}\label{62}
\frac {{\cal O}^3_n}{g_e}
= \frac {1}{6m_q}
\left [i\frac {g_V}{g_A} {\bf A}\cdot (\vep\times {\bf k})+\vsig\cdot ({\bf
A}\times (\vep\times {\bf k}))\right ]\frac 1{n!}\left (\frac
{{\bf k}\cdot {\bf q}}{3\alpha^2}\right )^n \nonumber \\
+\frac 1{18}\left [\frac {\omega_K\omega_{\gamma}}{\mu_s}\left (1+\frac
{\omega_{\gamma}}{2m_q}\right )\vsig \cdot \vep+\frac
{2}{\alpha^2}\vsig\cdot {\bf A}\vep\cdot {\bf q}\right ]\frac 1{(n-1)!}
  \left (\frac {{\bf k}\cdot {\bf q}}{3\alpha^2}\right )^{n-1}
\nonumber \\ +\frac {\omega_K\omega_\gamma}{54\alpha^2\mu_s}\frac 1{(n-2)!}
  \left (\frac {{\bf k}\cdot {\bf q}}{3
\alpha^2}\right )^{n-2}
\end{eqnarray}
and
\begin{eqnarray}\label{63}
\frac {{\cal O}^2_n}{g_S^\prime}
= \frac 1{6m_q} \left [ -ig_v^\prime {\bf A}\cdot (\vep \times
{\bf k})+g_a^\prime \vsig \cdot ({\bf A}\times (\vep\times {\bf k})\right ]
 \frac 1{n!}\left (\frac {-{\bf k}\cdot {\bf q}}{6\alpha^2}\right )^n
\nonumber \\
-\frac 1{36}\left [\frac {\omega_K\omega_{\gamma}}{\mu_s}
\left (1+g_a^\prime\frac {\omega_{\gamma}}{2m_q}\right )+\frac
{2\omega_{\gamma}}{\alpha^2} \vsig \cdot {\bf A}\vep\cdot {\bf q}\right ]
\frac 1{(n-1)!}\left (\frac {-{\bf k}\cdot {\bf q}}{6
\alpha^2}\right )^{n-1} \nonumber  \\
\frac {\omega_K\omega_\gamma}{216\alpha^2 \mu_s}\vsig \cdot {\bf
k}\vep\cdot {\bf q} \frac 1{(n-2)!}\left (\frac {-{\bf k}\cdot
{\bf q}}{6\alpha^2}\right )^{n-2}
\end{eqnarray}
where $g_S^\prime$ is the same as in Eq. \ref{38}, $g_v^\prime$ and
$g_a^\prime$ are given in Eqs. \ref{39} and \ref{40}, and $g_e$ is
2 for charged
kaons and $-1$ for the neutral kaons.  Generally, the resonances with
larger quantum number $n$ become important as the energy increases.
 Note that the amplitude ${\cal
O}_n^2$ generally differs from the amplitude ${\cal O}_n^3$ by a factor of
$\left (-\frac 12\right )^n$,  this shows that
the process that the incoming photon and outgoing kaon are
 absorbed and emitted by the same quark becomes more and more dominant
as the energy increases.  Furthermore,
the resonances with partial wave $l=n$ become dominant, of which the
isospin is 1/2 for $\gamma p\to K^+ \Lambda$ and 3/2 for $\gamma p\to
K\Sigma$.  Thus, we could use the mass and decay width of the high spin
states in Eq. \ref{43}; the resonance $G_{17}(2190)$ for the $n=3$
states and the resonance $H_{19}(2220)$ for the $n=4$ states
in $\gamma p\to K^+ \Lambda$, and the resonance $G_{37}(2200)$ for the
$n=3$ states and the resonance $H_{37}(2420)$ for n=4 states in $\gamma
p\to K\Sigma$.  Indeed, only the couplings for the high spin states
are strong enough to be seen experimentally, this is consistent with the
conclusions of the quark model

\subsection*{\bf 6. The Numerical evaluation}

It is straightforward to express every term in section 3, 4 and 5 in terms
of the CGLN amplitudes so that the differential cross section and various
polarization could be easily carried out.  However,
 it should be pointed out that the nonrelativistic
wavefunction is no longer adequate to describe the meson photoproduction
processes  because the relativistic effects become more and more important
as the energy increases.  A practical way to correct this shortcoming is
the introduction of the Lorentz boost factor\cite{lp72},  which was used
 in the calculations of the transition amplitudes as a
function of $Q^2$ in the electroproduction\cite{fh82}.  We shall adopt a
similar procedure in the meson-photoproduction for the transition amplitudes
as a function of the energies of initial and final state,
 the CGLN amplitudes become
\begin{equation}\label{64}
f_i({\bf k},{\bf q}) \to \frac {M_SM_N}{E_SE_N}f_i (\frac {M_N}{E_N}{\bf
k}, \frac {M_S}{E_S} {\bf q})
\end{equation}
where $i=1\dots 4$, and $\frac {M_N}{E_N}$ ($\frac {M_S}{E_S}$) is a Lorentz
boost factor for the initial (final) state.

The parameters used in our calculations have standard values except
the coupling constants $\alpha_K$ and $g_{K^*}$, which will be fitted to the
experimental data.  The quark masses are $m_q=0.34$ GeV for up and down
quarks and $m_s=0.55$ GeV for strange quarks.  Since the outgoing Kaons are
being considered point like particles in this calculation,
the constant $\alpha^2$ is chosen to be $0.2$ GeV$^2$ to take into account
part of the finite size effect, this is somewhat larger than that in the
calculation of the electromagnetic transition\cite{cko,simon}.  The masses
and the decay widths for the S-channel resonances are taken from the
recent particle data group\cite{pdg94}.

We present our calculation of the differential cross sections for
$\gamma p\to K^+ \Lambda$ at $E_{lab}=1.2$ in Fig. 1
and at $E_{lab}=1.4$ GeV in Fig. 2.  The coupling
constants  in this calculation are $\alpha_{KN\Lambda}\approx 4.0$
and $g_{K^*}\approx 0.4$.  The calculations of the differential cross
sections have been carried out from the threshold up to $E_{lab}=2.0$ GeV,
the results are in good agreement with the experimental
data\cite{ldata,bock94}.   There are some interesting
features that can be learnt here;  the differential cross section shows a
strong forward peaking around $E_{lab}=1.2$ GeV. As the energy
increases, the high partial wave resonances become increasingly important,
this leads to a backward peaking at high energies, which are dominated by
the high partial wave resonances.  Another important feature in the quark
model calculations is the form factors from the integrations of the
spatial wavefunction as well as the Lorentz boost factors; the total cross
section as a function of energies is shown in Fig. 3, the result is
consistent with the known data up to energy around 2 GeV in the laboratory
frame.  This has been a serious problem in the traditional hadronic
models\cite{as90,thom}

The processes $\gamma p\to K^+ \Sigma^0$ and $\gamma p\to K^0\Sigma^+$
can be calculated simultaneously in this approach; the difference between
the charged and neutral Kaon productions for the $\Sigma$ final states
is the absence of the seagull term and the Kaon exchange in the T-channel,
in the same time, the coupling constants $g_{K^*}$ in Eq. \ref{31}
should have opposite signs with approximately equal magnitudes
 in the two processes due to the magnetic dipole
transitions between $K^*$ and $K$\cite{godf}.  The U-channel and the
S-channel resonance contributions in the charged and neutral Kaon
productions for the $\Sigma$ final states can be related to each other by
the isospin couplings.  In Fig. 4 and 5, we show that the differential
cross sections for $\gamma p\to K^+ \Sigma^0$ at $E_{lab}=1.157$ and
$1.45$ GeV, where the data come  from the Bonn group\cite{bock94,bleck} and
Ref. \cite{anderson}.  The resulting coupling constants are
$\alpha_{K\Sigma P}=2.4$ and $g_{K^*}=-2.5$. The isospin 3/2 resonances, in
particular those belonging to ${\bf 56}$ multiplet, play a dominant role
for the $\Sigma$ final states. In the $E_{lab}=1.45$ GeV, the isospin 3/2
resonances $P_{31}(1910)$, $P_{33}(1920)$, $F_{35}(1950)$, and
$F_{37}(1905)$ contribute significantly, and the F-wave resonances
 lead to a larger differential cross section in the
backward angle.  In the Fig. 6, we show our results of the differential
cross section at $E_{lab}=1.15$ and $1.4$ GeV for $\gamma p\to
K^0\Sigma^+$.  Noticed that the seagull and s-channel nucleon pole
terms have opposite signs so that they tend to cancel each other for
$\gamma p\to K^+\Sigma^0$, while the cancelation does not exist for the
$\gamma p\to K^0\Sigma^+$ because of the absence of the Seagull term.  This
leads a larger total cross section in $\gamma p\to K^0\Sigma^+$,
and in the same time, the
differential cross sections become less forward peaked.   In Fig. 7, we
show that total cross section for both $\gamma p\to K^+\Sigma^0$ and
$K^0\Sigma^+$ comparing to the $\gamma p\to K^+\Sigma^0$ data\cite{bock94}.
Although the cross section for $\gamma p\to K^0\Sigma^+$ is larger than
that for $\gamma p\to K^+\Sigma^0$, this represents much more improved
calculations than the most hadronic models, which predict 10-100 times larger
cross section for the $K^0$ production than that for the $K^+$
production\cite{mart}.
The available  data\cite{braul} for neutral Kaon production, $\gamma p\to
K^0\Sigma^+$, are too poor  to test the theory. However, one could compare
the Kaon photoproduction for the $\Sigma$ final states to the $\pi$
photoproduction in the resonance $P_{33}(1232)$ region; in both cases, the
isospin 3/2 resonance dominant, therefore, the underline dynamics for the
two reactions are similar.  The cancelation among the Born terms in the
presence of the Seagull term leads to  a smaller cross section around the
peak of the resonance $P_{33}(1232)$ for the charge pion production,
$\gamma p\to \pi^+ n$,  although it dominates near the threshold,
while there is no cancelation
for the  neutral pion production, $\gamma p \to \pi^0 p$\cite{dt92}.
This shows that the neutral Kaon production data is crucial to provide
us insights into the dynamics of the meson-photoproductions.

No investigation of the Kaon as well as other meson photoproduction would
be complete without the calculation of the polarizations.  This can be
easily carried in our approach, because the expressions
 of every polarization in
terms of the CGLN amplitudes have been given in Refs. \cite{tabakin} and
 \cite{as90}. Here we only present our results for the recoil
asymmetry P defined as
\begin{equation}\label{80}
P=\frac
{{\frac {d\sigma}{d\Omega}}^{\uparrow}-{\frac {d\sigma}{d\Omega}}^{\downarrow}}
{{\frac {d\sigma}{d\Omega}}^{\uparrow}+{\frac {d\sigma}{d\Omega}}^{\downarrow}}
\end{equation}
where $\uparrow (\downarrow)$ corresponds to the spin of the final state
parallel
(antiparallel) to the direction of the vector ${\bf k}\times {\bf q}$.
  The recoil asymmetry P is very important since it provides us a direct
probe to the structure of the S-channel resonances in our approach.  As the
decay width for a resonance become large, it is important to treat it as a
function of the momentum of the final decay particles;  the decay
width reaches its peak as the center mass energy equals to its mass, and
decreases quickly if the total energy of the system moves away from the
mass of the resonances.  Therefore, it is only necessary to introduce the
decay width for the S-channel resonances whose masses are higher than the
threshold of the Kaon photoproduction.   This shows that the imaginary part
of the transition amplitudes only depends on the structure of the S-channel
resonances above threshold, and there could be an overall phase factor for
these resonances.  In Fig. 8, we show the recoil asymmetry P as a function
of $E_{lab}$ at 90 degrees for both $\gamma p\to$ $K^+\Lambda$ and
$K^+\Sigma^0$, the asymmetry for the $\gamma p\to K^+\Sigma^0$ is small and
positive comparing to
the negative  $\gamma p\to K^+\Lambda$ polarization, and
this of course is in agreement with the experimental
data\cite{bock94,thom}.  The physics for the opposite signs between the
$\Lambda$ and $\Sigma$ final states is the dominance of the isospin 3/2
resonances in $\gamma p\to K^+\Sigma^0$.
Perhaps more interesting  feature of this calculation is the changing sign
of the recoil asymmetry for $\gamma p\to K^+\Lambda$ in the backward angle
as the energy increases.  In Fig. 9, we show the asymmetry P for $\gamma p\to
K^+\Lambda$ as a function of the scattering angle at $E_{lab}=1.157$ and
$1.4$ GeV, the asymmetry P is negative at $E_{lab}=1.1$ GeV and becomes
positive in the backward angle at $1.4$ GeV.  Indeed, the recent data from
Bonn group\cite{bock94} show some hints that this might be indeed the case.
This feature is new and has not been shown in early
calculations\cite{as90}, the reason for the changing signs is the higher
partial wave resonances become more important as the energy increases.

\subsection*{\bf 7. Future Work and Summary}

Our calculations show that the chiral quark model presents a much better
framework to understand the reaction mechanism of the Kaon
photoproductions than many traditional hadronic models.
In the same time, it also raises some interesting issues
that need to be
further studied.  First, the relation between the chiral quark model
and the Quark Pair Creation (QPC) Model\cite{LY} which has been quite
successful phenomenologically in studying the strong decays of
hadrons\cite{robert}.
Thus, it is natural to calculate the meson-photoproduction in the
QPC model, which has been done by Kumar and Onley\cite{kumar}.
  The problem in their
approach is that the QPC model alone does not lead to the low energy theorem
near the threshold.  This can be seen in the Seagul term that dominates the
low energy behaviour; the simple gauge transformation in the QPC leads to
a Seagul term that is equivalent to the process $\gamma \to s\bar {s}$ inside
nucleons.  However, the chiral transformation tells us that there is
another Seagul
term that is equivalent to the process $\gamma u (d) \to u (d)$ for
$K^+$($K^0$) production, which is not present in the simple QPC model.
The relation between the two terms is determined by
the chiral transformation; thus the total Seagul term has the structure;
\begin{equation}\label{70}
{\cal M}_{sea}=e_{u(d)}R(\gamma u(d)\to u(d))-e_sR(\gamma \to s\bar {s})
\end{equation}
where $e_{u(d)}$ is the charge of the up quark for the $K^+$ production and
the down quarks for the $K^0$ production, $e_s$ is the charge of strange quarks
 and $R$ represents the spatial
integral for each process.  If the finite size of the emitting Kaon is
being taken into account, the spatial integrals for the two processes are not
equal;
\begin{equation}\label{71}
R(\gamma u(d)\to u(d))=N(\alpha^2,\beta^2)e^{-\frac 1{1+\frac
{2\alpha^2}{3\beta^2}}\left [\frac {1+\frac
{\alpha^2}{\beta^2}}{6\alpha^2}({\bf k}-{\bf q})^2+\frac {\alpha^2}{
18\beta^4}\left ({\bf k}^2-\frac 12{\bf q}^2\right )\right ]}
\end{equation}
and
\begin{equation}\label{72}
R(\gamma \to s\bar {s})=N(\alpha^2,\beta^2)e^{-\frac 1{1+\frac
{2\alpha^2}{3\beta^2}}\left [\frac 1{6\alpha^2}({\bf k}-{\bf q})^2
+\frac {5\alpha^2}{
9\beta^2}\left (\frac 23{\bf k}^2+\frac 14\left (5+4\frac
{\alpha^2}{\beta^2}\right ){\bf q}^2\right )\right ]},
\end{equation}
where $\beta^2$ and $\alpha^2$ are the harmonic oscillator strengths for
the meson and baryon wavefunctions, and $N(\alpha^2,\beta^2)$ is the
normalization factor,  which is the same for both processes.   The
consequences for the finite size mesons are that
there are additional terms proportional to
on $q^2$ and $k^2$ which makes the the Seagull term decreases faster as
the momentum ${\bf k}$ and ${\bf q}$ is increases, and it also generates an
correction at order $\left (\frac {m_{\pi}}{M_N}\right )^2$ to the low
energy theorem near the threshold\cite{zpli94} for both charge and neutral
 pion photoproductions.  Moreover, a nonzero contribution from the
Seagull term could affect the result of $\gamma p\to K^0\Sigma^+$
significantly at higher energies.
  This has not been discussed in the literature,
and certainly deserves more study.

Second; the relativistic effects in the transition
operator\cite{close} should be included. Although the nonrelativistic
transition operator used in this calculation is sufficient to reproduce
the low energy theorem in threshold region, the relativistic corrections
in the electromagnetic transition operator is important.  In particular,
the spin-orbital and so called nonadditive terms are required to
derive the low energy theorem in the Compton scattering\cite{zpli93} and
Drell Hearn Gerasimov sum rule\cite{zpli92}, and they also breaks
Moorhouse selection rule so that the resonance $D_{13}(1700)$ would
also contribute to the Kaon productions.  Furthermore, the more
consistent baryon wavefunction\cite{IK79}
should be used in the calculation.

Third; the role of gluonic degrees of freedom  in the baryon wavefunction
should be investigated.  Our  calculation shows a small contribution from
the Roper resonance if it is a radial excitation of the nucleon, however, a
hybrid Roper resonance would increase the contribution from the Roper
resonance as well as the resonance $P_{33}(1600)$ significantly, the
consequences of the hybrid Roper resonance and $P_{33}(1600)$ in the Kaon
photoproductions remain to be studied.

Because these effects remain to be studied, the parameters, in particular
the kaon coupling constants $\alpha_K$, obtained in our calculations are by
no means final, nor  they are consistent with the measurements in
Kaon-Nucleon scattering\cite{martin}.
Our  investigation does provide
a framework to give a description of the Kaon-photoproductions
that is consistent with the Kaon-nucleon scattering.  Moreover, this approach
makes it possible to provide a unified description for $\pi$, $\eta$ and
kaon photoproductions. The extension to the $\eta$ and $\pi$ photoproduction
is in progress, and will be published elsewhere.

\subsection*{Acknowledgement}
The author would like to thank L. Kisslinger for his encouragement and
discussions, and D. Onley, R. Shumacher for providing me the
update information on the Kaon photoproductions and useful discussions.
Discussions with F. Tabakin, C. Bennhold, B. Keister, S. Dytman, and C. Benesh
 are also gratefully acknowledged.
This work is supported by U.S. National Science Foundation grant
PHY-9023586.

\newpage

\subsection*{Figure Caption}
\begin{enumerate}
\item The differential cross section for $\gamma p\to K^+\Lambda$ at
$E_{lab}=1.2$ GeV, it has a $10^{-30}cm^2/sr$ unit.
\item The same as Fig. 1 at $E_{lab}=1.4$ GeV.
\item The total cross section for $\gamma p\to K^+\Lambda$ as a function
of the energy in the laboratory frame $E_{lab}$, it has a $10^{-30} cm^2$
unit.
\item The differential cross section
for $\gamma p\to K^+\Sigma^0$ at $E_{lab}=1.157$ GeV, it has a $10^{-30}
cm^2/sr$ unit.
\item The same as Fig. 4 at $E_{lab}=1.45$ GeV.
\item The differential cross sections for $\gamma p\to K^0\Sigma^+$,
the solid and dashed lines correspond to $E_{lab}=1.4$ and $1.15$ GeV
respectively.
\item The total cross sections as a function of $E_{lab}$, the data and
solid line correspond to the process $\gamma p\to K^+\Sigma^0$, while
the dashed line is the prediction for $\gamma p\to K^0\Sigma^+$.
\item The recoil asymmetry $P$ at 90 degrees as a function of
$E_{lab}$; the solid line and data
correspond to the process $\gamma p\to K^+\Lambda$, and the dashed line
represents  $\gamma p\to K^+\Sigma^0$.
\item The asymmetry $P$ as a function of scattering angle for $\gamma p
\to K^+\Lambda$ at $E_{lab}=1.1$ (solid line) and $1.4$ (dashed line) GeV
comparing to the data at $E_{lab}=1.1$ GeV.
\end{enumerate}
\end{document}